\documentclass{pspum-l}

\usepackage{}
\input xy
\xyoption{all}

\newtheorem{theorem}{Theorem}[section]

\theoremstyle{definition}

\theoremstyle{remark}

\numberwithin{equation}{section}

\begin{document}

\title{A few recent developments in 2d (2,2) and (0,2) theories}


\author[E. Sharpe]{Eric Sharpe}
\address{Department of Physics, Virginia Tech}
\curraddr{Department of Physics, Virginia Tech}
\email{ersharpe@vt.edu}
\thanks{ES was partially supported by NSF grants PHY-1068725, PHY-1417410.}



\date{}

\begin{abstract}
In this note we summarize a few of the many
recent developments in two-dimensional quantum field theories.
We begin with a review of the current state of quantum sheaf
cohomology, a heterotic analogue of quantum cohomology.
We then turn to dualities:
we outline the current status of (0,2) mirror symmetry,
and then outline recent work on two-dimensional gauge dualities.
In particular, we 
describe how many
two-dimensional gauge dualities in both (2,2) and (0,2) supersymmetric
gauge theories can be understood simply as different presentations of the
same infrared (IR) geometry.  We then discuss (not necessarily supersymmetric)
two-dimensional nonabelian gauge
theories in which a subgroup of the gauge group acts trivially on 
massless matter.  We describe how
these theories `decompose' into disjoint unions of 
other theories indexed by discrete theta angles, a fact
which in other contexts
has proven to have implications for interpretations of
certain gauged linear sigma models (GLSMs) 
and for Gromov-Witten invariants of stacks.
We conclude with a discussion of recent developments in infinitesimal
moduli of heterotic compactifications.
\end{abstract}

\maketitle

\section{Introduction}

Over the last half dozen years, there
has been a tremendous amount of progress
in gauged linear sigma models (GLSMs) and perturbative string compactifications.
A few examples include, but are not limited to:
\begin{itemize}
\item Nonperturbative realizations of geometry in GLSMs
\cite{cdhps,meron,hori-tong},
\item Perturbative realizations of Pfaffians \cite{hori2,hori-knapp,jsw,jklmr},
\item Non-birational GLSM phases, and physical realizations of homological
projective duality \cite{Ballard:2013fxa,Ballard2,cdhps,meron,halpern-leistner,hori-tong,kuz3,kuz1,kuz2},
\item Examples of closed strings on noncommutative resolutions
\cite{add-ss,cdhps,ncgw},
\item Localization techniques, yielding new Gromov-Witten and elliptic
genus computations, the role of Gamma classes, and much more
(see {\it e.g.} \cite{bc1,beht2,beht,doroudetal,gg,jklmr2} for a few 
references),
\item Heterotic strings:  nonperturbative corrections, 2d dualitites,
and non-K\"ahler moduli \cite{ade,ags,hetstx,shifman2,delaOssa:2014cia,dgks1,dgks2,dgks-sm11,guffinrev,gk1,gjs,ks1,lin-wu-yau,mcoristrev,mm-half,mm-summing,mss,mp-02mirror,ms-nk,es-snowbird,es-other,qsc-grenoble,tan2,tan1}.
\end{itemize}
This talk will largely, though not exclusively,
focus on heterotic strings.  We
will survey some of the results
in two-dimensional (0,2) theories over the last six years or so,
describing both new results as well as outlining some older results to
help provide background and context.

We begin in section~\ref{sect:qsc} with a brief review of the current state of 
the art in quantum sheaf cohomology.  In section~\ref{sect:02mirror} we
give a brief status report on (0,2) mirror symmetry.
In section~\ref{sect:duals} we discuss recent progress in 
two-dimensional gauge dualities in theories with (2,2) and (0,2)
supersymmetry.  We discuss how a number of Seiberg-like dualities can
be understood simply as different presentations of the same IR geometry,
and use this to predict additional dualities.
In section~\ref{sect:decomp} we turn to a different gauge duality,
one that applies to both supersymmetric and nonsupersymmetric theories in
two dimensions.  Specifically, in two-dimensional gauge theories in which
a finite subgroup of the gauge group acts trivially on the matter, the
theory `decomposes' into a disjoint union of theories.  In nonabelian gauge
theories, the various components are labelled by different discrete theta
angles.
Finally, in section~\ref{sect:moduli} we discuss current progress in
infinitesimal moduli in heterotic compactifications, specifically,
recent developments in understanding moduli
in both Calabi-Yau and also non-K\"ahler heterotic compactifications.

\section{Review of quantum sheaf cohomology}
\label{sect:qsc}

Quantum sheaf cohomology is the heterotic string analogue of quantum
cohomology.  Whereas ordinary quantum cohomology is defined by a space,
quantum sheaf cohomology is defined by a space together with a bundle.
Specifically, quantum sheaf cohomology is defined by a complex manifold
$X$
together with a holomorphic vector bundle ${\mathcal E} \rightarrow X$
(often called the `gauge bundle'),
satisfying the conditions
\begin{displaymath}
{\rm ch}_2({\mathcal E}) \: = \: {\rm ch}_2(TX), \: \: \:
\det {\mathcal E}^* \: \cong \: K_X .
\end{displaymath}
Briefly, whereas ordinary quantum cohomology is defined by intersection
theory on a moduli space of curves, quantum sheaf cohomology is defined
by sheaf cohomology (of sheaves induced by ${\mathcal E})$) over a moduli
space of curves.  In the special case that ${\mathcal E} = TX$,
the quantum sheaf cohomology ring should match the ordinary
quantum cohomology ring.
See for example \cite{dgks1,dgks2,dgks-sm11,mcoristrev,mss} 
for a few recent discussions.  We shall give here a brief summary oriented
more nearly towards physicists; see for example \cite{dgks-sm11} for
a longer summary oriented towards mathematicians.

In heterotic string compactifications, quantum sheaf cohomology encodes 
nonperturbative corrections to charged matter couplings.  For example,
for a heterotic compactification on a Calabi-Yau three-fold $X$ with gauge
bundle given by the tangent bundle (known as the standard embedding, 
or as the (2,2) locus, as in this case (0,2) supersymmetry is enhanced to
(2,2)),
the low-energy theory has an $E_6$ gauge symmetry and matter charged
under the ${\bf \overline{27}}$, counted by $H^{1,1}(X)$. 
The nonperturbative corrections to the ${\bf \overline{27}}^3$ couplings
are encoded in Gromov-Witten invariants \cite{candelas-mirror}
and computed by the A model
topological field theory \cite{witten-tft}.

If we now deform the gauge bundle so that it is no longer the tangent
bundle, then the ${\bf \overline{27}}^3$ couplings will still receive
nonperturbative corrections, but those corrections are no longer computed
by Gromov-Witten invariants or the A model.  Instead, the nonperturbative
corrections are encoded in quantum sheaf cohomology.
In this more general context,
mathematical Gromov-Witten computational tricks no longer seem to apply,
and there is no known analogue of periods or Picard-Fuchs equations.
New methods are needed, and a few new techniques have been developed,
which will be outlined here.

Before working through details, let us give a simple example.
Recall the ordinary quantum cohomology ring of
${\mathbb P}^n$ is given by
\begin{displaymath}
{\mathbb C}[x]/(x^{n+1} - q) .
\end{displaymath}
When $q \rightarrow 0$, this becomes the classical cohomology ring of
${\mathbb P}^n$, hence the name.  Now, to compare, the quantum sheaf
cohomology ring of ${\mathbb P}^n \times {\mathbb P}^n$ with bundle
${\mathcal E} \rightarrow {\mathbb P}^n \times {\mathbb P}^n$ defined by
\begin{displaymath}
0 \: \longrightarrow \: {\mathcal O} \oplus {\mathcal O} \: 
\stackrel{*}{\longrightarrow} \: 
{\mathcal O}(1,0)^{n+1} \oplus {\mathcal O}(0,1)^{n+1}
\: \longrightarrow \: {\mathcal E} \: \longrightarrow \: 0 ,
\end{displaymath}
where
\begin{displaymath}
* \: = \: \left[ \begin{array}{cc}
A x & B x \\
C \tilde{x} & D \tilde{x}
\end{array} \right] 
\end{displaymath}
($x, \tilde{x}$ vectors of homogeneous coordinates on the two
${\mathbb P}^n$'s, $A, B, C, D$ a set of four
$(n+1)\times (n+1)$ constant matrices encoding a deformation of the
tangent bundle), 
is given by
\begin{displaymath}
{\mathbb C}[x,y] / \left( \det( Ax + By) - q_1, \:
\det(Cx + D y) - q_2 \right) .
\end{displaymath}

Note that in the special case that $A=D=I$, $B=C=0$, the bundle ${\mathcal E}$
coincides with the tangent bundle of ${\mathbb P}^n \times {\mathbb P}^n$,
and in this case, the quantum sheaf cohomology ring above reduces to
\begin{displaymath}
{\mathbb C}[x,y]/(x^{n+1}-q_1, y^{n+1}-q_2) ,
\end{displaymath}
which is precisely the ordinary quantum cohomology ring of 
${\mathbb P}^n \times {\mathbb P}^n$.  This is as expected:
as mentioned earlier,  
when ${\mathcal E} = 
TX$, quantum sheaf cohomology reduces to ordinary quantum cohomology.

Ordinary quantum cohomology can be understood physically as the ring of local
operators, known as the OPE ring, of
the A model topological field theory in two dimensions.  That topological
field theory is obtained by twisting a (2,2) nonlinear sigma model along
a vector $U(1)$ symmetry.  
In a (0,2) nonlinear sigma model, if $\det {\mathcal E}^*
\cong K_X$, then there is a nonanomalous $U(1)$ symmetry one can twist along,
which reduces to the vector $U(1)$ symmetry on the (2,2) locus.  If we twist
along that nonanomalous $U(1)$, the result is a pseudo-topological field
theory known as the A/2 model.  Quantum sheaf cohomology is the OPE ring
of the A/2 model.  (There is also a pseudo-topological analogue of the
B model, known as the B/2  model, but in this lecture we shall focus
on the A/2 model.)

To be consistent, the ring products must close into the ring, but this is
not {\it a priori} automatic in these quantum field theories, 
as in principle the products might generate
local operators which are not elements of the (pseudo-)topological
field theory.  In the case of (2,2) supersymmetry, this closure of the OPE
ring was
argued in {\it e.g.} \cite{lvw}.  Closure in (0,2) theories is also
possible -- closure does not require (2,2) supersymmetry,
but can be accomplished under weaker conditions.
This was studied in detail in \cite{ade}.  For example, for a (0,2)
SCFT, one
can use a combination of worldsheet conformal invariance and the right-moving
$N=2$ algebra to argue closure of the OPE ring on patches on the moduli space.

The local operators in the A model, the additive part of the OPE ring,
are BRST-closed states of the form
\begin{displaymath}
b_{i_1 \cdots i_p \overline{\imath}_1 \cdots \overline{\imath}_q}
\chi^{\overline{\imath}_1} \cdots \chi^{\overline{\imath}_q}
\chi^{i_1} \cdots \chi^{i_p} ,
\end{displaymath}
which are identified with closed differential forms representing
$H^{p,q}(X)$.  The analogous operators in the A/2 model are right-BRST-closed
states of the form
\begin{displaymath}
b_{\overline{\imath}_1 \cdots \overline{\imath}_q a_1 \cdots a_p}
\psi_+^{\overline{\imath}_1} \cdots \psi_+^{\overline{\imath}_q}
\lambda_-^{a_1} \cdots \lambda_-^{a_p} ,
\end{displaymath}
which are identified with closed bundle-valued differential forms
representing elements of $H^q(X, \wedge^p {\mathcal E}^*)$.
On the (2,2) locus, where ${\mathcal E} = TX$,
the A/2 model reduces to the A model, which in
operators follows from the statement
\begin{displaymath}
H^q(X, \wedge^p T^*X) \: = \: H^{p,q}(X) .
\end{displaymath}

At a purely schematic level, we can understand correlation functions as
follows.  Classically, in the A model, correlation functions are of the
form
\begin{displaymath}
\langle {\mathcal O}_1 \cdots {\mathcal O}_n \rangle \: = \:
\int_X \omega_1 \wedge \cdots \wedge \omega_n \: = \:
\int_X \left( \mbox{top-form} \right) ,
\end{displaymath}
where $\omega_i \in H^{p_i,q_i}(X)$.
In the A/2 model, classical contributions to correlation functions 
are of the form
\begin{displaymath}
\langle {\mathcal O}_1 \cdots {\mathcal O}_n \rangle \: = \:
\int_X \omega_1 \wedge \cdots \wedge \omega_n ,
\end{displaymath}
where $\omega_i \in H^{q_i}(X, \wedge^{p_i} {\mathcal E}^*)$.
Now, 
\begin{displaymath}
\omega_1 \wedge \cdots \wedge \omega_n \: \in \:
H^{\rm top}(X, \wedge^{\rm top} {\mathcal E}^*)
\: = \: H^{\rm top}(X, K_X) ,
\end{displaymath}
using the anomaly constraint $\det {\mathcal E}^* \cong K_X$.  Thus, again
we have a top-form, and so the correlation function yields a number.

In passing, note that the number one gets above depends upon a particular
choice of an isomorphism $\det {\mathcal E}^* \cong K_X$.  To uniquely
define the A/2 theory, one must pick a particular isomorphism, which
is a reflection of properties of the corresponding physical heterotic
worldsheet theory.  Moreover,
as one moves on the moduli space of bundles or complex or K\"ahler
structures, that isomorphism may change, so these correlation functions
should be understood as sections of bundles over such moduli spaces.
Technically, this is closely related to the realization of
the Bagger-Witten line bundle in four-dimensional $N=1$ supergravity
\cite{bagger-ed} on the worldsheet \cite{dist-trieste,ps1}, as the action
of the global $U(1)$ in the worldsheet $N=2$ algebra on the spectral
flow operator.
(The original Bagger-Witten paper \cite{bagger-ed} assumed that the
SCFT moduli space was a smooth manifold; see for example 
\cite{dist-me,simeon-me} for modern generalizations 
to the case of moduli stacks.)

Correlation functions as outlined above define functions on
spaces of sheaf cohomology groups.  Now, we are interested in
the relations amongst products of those sheaf cohomology groups, and those
relations emerge as kernels of the (correlation) functions.

Let us consider a concrete example, namely the classical sheaf
cohomology of ${\mathbb P}^1 \times {\mathbb P}^1$ with bundle
${\mathcal E}$ given by a deformation of the tangent bundle, defined as
\begin{equation}  \label{eq:e-defn}
0 \: \longrightarrow \: W^* \otimes {\mathcal O} \:
\stackrel{*}{\longrightarrow} \: 
{\mathcal O}(1,0)^2 \oplus {\mathcal O}(0,1)^2 \: \longrightarrow \:
{\mathcal E} \: \longrightarrow \: 0 ,
\end{equation}
where $W \cong {\mathbb C}^2$,
\begin{displaymath}
* \: = \: \left[ \begin{array}{cc}
A x & B x \\
C \tilde{x} & D \tilde{x} \end{array} \right],
\end{displaymath}
$x, \tilde{x}$ vectors of homogeneous coordinates on the two ${\mathbb P}^1$'s,
and $A, B, C, D$ four $2 \times 2$ constant matrices encoding the tangent
bundle deformation.

We will focus on operators counted by 
\begin{displaymath}
H^1({\mathcal E}^*) \: = \: H^0(W \otimes {\mathcal O}) \: = \: W .
\end{displaymath}
An $n$-point correlation function is then a map
\begin{displaymath}
{\rm Sym}^n H^1({\mathcal E}^*) \: \left( \, = \: {\rm Sym}^n W \, \right) \:
\longrightarrow \: H^n\left( \wedge^n {\mathcal E}^* \right) .
\end{displaymath}
The kernel of this map defines the classical sheaf cohomology ring
relations, which we shall compute.

Since ${\mathcal E}$ is rank two, we will consider products of two elements
of $H^1({\mathcal E}^*) = W$, {\it i.e.} a map
\begin{displaymath}
H^0\left( {\rm Sym}^2 W \otimes {\mathcal O} \right) \: \longrightarrow \:
H^2(\wedge^2 {\mathcal E}^* ) .
\end{displaymath}
This map is implicitly encoded in the resolution
\begin{equation}  \label{eq:e-resn}
0 \: \longrightarrow \: \wedge^2 {\mathcal E}^* \: \longrightarrow \:
\wedge^2 Z \: \longrightarrow \: Z \otimes W \: \longrightarrow \:
{\rm Sym}^2 W \otimes {\mathcal O} \: \longrightarrow \:
0 ,
\end{equation}
determined by the definition~(\ref{eq:e-defn}), where
\begin{displaymath}
Z \: \equiv \:
{\mathcal O}(-1,0)^2 \oplus {\mathcal O}(0,-1)^2 .
\end{displaymath}
We break the resolution~(\ref{eq:e-resn}) into a pair of short exact
sequences:
\begin{equation}  \label{eq:break1}
0 \: \longrightarrow \: \wedge^2 {\mathcal E}^* \: \longrightarrow \:
\wedge^2 Z \: \longrightarrow \: S_1 \: \longrightarrow \: 0 ,
\end{equation}
\begin{equation}  \label{eq:break2}
0 \: \longrightarrow \: S_1 \: \longrightarrow \: Z \otimes W \:
\longrightarrow \: {\rm Sym}^2 W \otimes {\mathcal O} \: \longrightarrow \: 0 ,
\end{equation}
(which define $S_1$).

The second sequence~(\ref{eq:break2}) induces
\begin{displaymath}
H^0(Z \otimes W) \: \longrightarrow \: H^0({\rm Sym}^2 W \otimes {\mathcal O})
\: \stackrel{\delta}{\longrightarrow} \: H^1(S_1) \: \longrightarrow \:
H^1(Z \otimes W) .
\end{displaymath}
Since $Z$ is a sum of ${\mathcal O}(-1,0)$'s and ${\mathcal O}(0,-1)$'s,
\begin{displaymath}
H^0(Z \otimes W) \: = \: 0 \: = \:
H^1(Z \otimes W) ,
\end{displaymath}
hence
the coboundary map
\begin{displaymath}
\delta: \: H^0({\rm Sym}^2 W \otimes {\mathcal O})
\: \stackrel{\sim}{\longrightarrow} \: H^1(S_1)
\end{displaymath}
is an isomorphism.

The first sequence~(\ref{eq:break1}) induces
\begin{displaymath}
H^1(\wedge^2 Z) \: \longrightarrow \: H^1(S_1) \: 
\stackrel{\delta}{\longrightarrow} \:
H^2(\wedge^2 {\mathcal E}^*) \: \longrightarrow \:
H^2(\wedge^2 Z) .
\end{displaymath}
The last term vanishes, but $H^1(\wedge^2 Z) \cong {\mathbb C}^2$, hence
the coboundary map
\begin{displaymath}
\delta: \: H^1(S_1) \: \longrightarrow \: H^2(\wedge^2 {\mathcal E}^*)
\end{displaymath}
has a two-dimensional kernel.

The composition of these two coboundary maps is our designed two-point
correlation function
\begin{displaymath}
H^0({\rm Sym}^2 W \otimes {\mathcal O}) \: 
\stackrel{\delta, \sim}{\longrightarrow} \:
H^1(S_1) \: \stackrel{\delta}{\longrightarrow} \:
H^2(\wedge^2 {\mathcal E}^*) .
\end{displaymath}
The right $\delta$ has a two-dimensional kernel, which one can show is
generated by
\begin{displaymath}
\det(A \psi + B \tilde{\psi}), \: \: \:
\det(C \psi + D \tilde{\psi}) ,
\end{displaymath}
where $A, B, C, D$ are four matrices defining the deformation ${\mathcal E}$,
and $\psi, \tilde{\psi}$ correspond to elements of a basis for $W$.

Putting this together, we get that the classical sheaf cohmology ring is
\begin{displaymath}
{\mathbb C}[\psi,\tilde{\psi}] / \left(
\det(A \psi + B \tilde{\psi}), \det(C \psi + D \tilde{\psi}) \right) .
\end{displaymath}

So far we have discussed classical physics.  Instanton sectors have the
same general form, except that $X$ is replaced by a moduli space $M$ of curves,
and ${\mathcal E}$ is replaced by an induced sheaf\footnote{
If there are vector zero modes (`excess' intersection in the (2,2) case),
then this story is more complicated -- for example, there is a second induced
sheaf, and one must utilize four-fermi terms in the action.
For simplicity, for the purposes of this outline,
we shall focus on the simpler case of no vector zero modes.
} ${\mathcal F}$
over the moduli
space $M$.  Broadly speaking, the moduli space $M$ must be compactified,
and ${\mathcal F}$ extended over the compactification divisor.
The anomaly conditions
\begin{displaymath}
{\rm ch}_2({\mathcal E}) \: = \: {\rm ch}_2(TX), \: \: \:
\det {\mathcal E}^* \: \cong \: K_X
\end{displaymath}
imply that
\begin{displaymath}
\det {\mathcal F}^* \: \cong \: K_M ,
\end{displaymath}
which is needed for the correlation functions to yield numbers.

Within any one instanton sector, in general terms one can follow the same 
method just outlined.
In the case of the example just outlined, it can be shown that in a sector
of instanton degree $(a,b)$, the `classical' ring in that sector is of
the form
\begin{displaymath}
{\rm Sym}^{\bullet} W / (Q^{a+1}, \tilde{Q}^{b+1} ) ,
\end{displaymath}
where
\begin{displaymath}
Q \: = \: \det(A \psi + B \tilde{\psi}), \: \: \:
\tilde{Q} \: = \: \det(C \psi + D \tilde{\psi}) .
\end{displaymath}

Now, OPE's can relate correlation functions in different instanton
degrees, and so should map ideals to ideals.  To be compatible with
the ideals above,
\begin{displaymath}
\langle {\mathcal O} \rangle_{a,b} \: = \:
q^{a'-a} \tilde{q}^{b'-b} \langle {\mathcal O} Q^{a'-a}
\tilde{Q}^{b'-b} \rangle_{a',b'}
\end{displaymath}
for some constants $q, \tilde{q}$.  As a result of the relations above,
we can read off the OPE's
\begin{displaymath}
Q \: = \: q, \: \: \:
\tilde{Q} \: = \: \tilde{q} ,
\end{displaymath}
which are the quantum sheaf cohomology relations.

More generally \cite{dgks1,dgks2,mm-summing},
for any toric variety, and any deformation ${\mathcal E}$ of its 
tangent bundle defined in the form
\begin{displaymath}
0 \: \longrightarrow \: W^* \otimes {\mathcal O} \: 
\stackrel{*}{\longrightarrow} \:
\underbrace{ \oplus_i {\mathcal O}(\vec{q}_i) }_{Z^*} \: \longrightarrow \:
{\mathcal E} \: \longrightarrow \: 0 ,
\end{displaymath}
the chiral ring
is
\begin{displaymath}
\prod_{\alpha} \left( \det M_{(\alpha)} \right)^{Q^a_{\alpha}} \: = \:
q_a ,
\end{displaymath}
where the $M_{(\alpha)}$'s are matrices of chiral operators constructed
from the map $*$.

So far we have outlined mathematical computations of quantum sheaf cohomology,
but there also exist methods based on gauged linear sigma models (GLSMs):
\begin{itemize}
\item Ordinary quantum cohomology is computed from (2,2) GLSMs in
\cite{mp1},
\item Quantum sheaf cohomology is computed from (0,2) GLSMs in
\cite{mm-half,mm-summing}.
\end{itemize}
Briefly, for the (0,2) case, one computes quantum corrections to the
effective action of the form
\begin{displaymath}
L_{\rm eff} \: = \: \int d \theta^+ \sum_a \Upsilon_a \log\left(
\prod_{\alpha} (\det M_{(\alpha)} )^{Q^a_{\alpha}} / q_a \right) ,
\end{displaymath}
from which one derives the conditions for vacua
\begin{displaymath}
\prod_{\alpha} \left( \det M_{(\alpha)} \right)^{Q^a_{\alpha}} \: = \:
q_a .
\end{displaymath}
These are the quantum sheaf cohomology relations, and those derived in
\cite{dgks1,dgks2} match these.

The current state of the art in quantum sheaf cohomology are computations
on toric varieties.  
Our goal is to eventually perform these computations on compact Calabi-Yau
manifolds, and as an intermediate step, we are currently studying 
Grassmannians.  

Briefly, we need better computational methods.  Conventional Gromov-Witten
tricks seem to revolve around the idea that the A model is independent of
complex structure, which is not necessarily true for the A/2 model.
That said, it has been argued \cite{mm-summing} that the A/2 model is
independent of some moduli.  Despite attempts to check \cite{Garavuso:2013zoa},
however, this is still not perfectly well-understood.

\section{(0,2) mirror symmetry}
\label{sect:02mirror}

Let us begin our discussion of dualities with a review of progress on
a conjectured generalization of mirror symmetry, known as (0,2) mirror
symmetry.  

Now, ordinary mirror symmetry, in its most basic form, is a relation
between Calabi-Yau manifolds, ultimately because a (2,2) supersymmetric
nonlinear sigma model is defined by a manifold.
Nonlinear sigma models with (0,2) supersymmetry are defined by a space $X$
together with a holomorphic vector bundle ${\mathcal E} \rightarrow X$
satisfying certain consistency conditions discussed earlier, 
so (0,2) mirror symmetry,
in its most basic form, is a statement about complex manifolds
together with holomorphic vector bundles.

In this language, a prototypical\footnote{
Described here is the most basic incarnation of (0,2) mirror symmetry.
For example, ordinary mirrors are sometimes given by
Landau-Ginzburg models instead of spaces, and there are analogous statements
in the (0,2) case.  For simplicity, we focus on prototypical incarnations
in which both sides of the mirror relation are defined by spaces (and
bundles).
} (0,2) mirror is defined by a space $Y$
with holomorphic vector bundle ${\mathcal F} \rightarrow Y$, such that
\begin{eqnarray*}
{\rm dim}\, X & = & {\rm dim}\, Y, \\
{\rm rank}\, {\mathcal E} & = & {\rm rank}\, {\mathcal F}, \\
{\rm A/2}(X, {\mathcal E}) & = & {\rm B/2}(Y, {\mathcal F}), \\
H^p(X, \wedge^q {\mathcal E}^*) & = &
H^p(Y, \wedge^q {\mathcal F}), \\
({\rm moduli}) & = & ({\rm moduli}).
\end{eqnarray*}
In the special case that ${\mathcal E} = TX$, (0,2) mirror symmetry should
reduce to ordinary mirror symmetry.

Some of the first significant evidence for (0,2) mirror symmetry was
numerical:  the authors of \cite{bsw1} wrote a computer program to scan a large
number of examples and compute pertinent sheaf cohomology groups.
The resulting data set was mostly invariant under the exchange of sheaf
cohomology groups outlined above, giving a satisfying albeit limited test
of the existence of (0,2) mirrors.

In \cite{ss1}, the Greene-Plesser orbifold construction \cite{gp1} was extended
to (0,2) models.  This construction (and its (0,2) generalization) creates
mirrors to Fermat-type Calabi-Yau hypersurfaces and complete
intersections, as resolutions of certain orbifolds of the original
hypersurface or complete intersection.  Because the construction can
be understood as utilizing symmetries of what are called `Gepner models'
(see {\it e.g.} \cite{gepner1}), 
the fact that the SCFT's match is automatic, and so one can build what are
necessarily examples of (0,2) mirrors.  Unfortunately,
this construction does not generate families of mirrors, only isolated
examples.

In another more recent development, the Hori-Vafa-Morrison-Plesser-style
GLSM duality picture of mirror symmetry \cite{hv1,mp0} was repeated for (0,2)
theories in \cite{abs}.  Unfortunately, unlike the case of ordinary
mirror symmetry, understanding duality in (0,2) GLSMs requires additional
input beyond the machinery of \cite{hv1,mp0}.

More recently, a promising approach was discussed in
\cite{mp-02mirror}, generalizing Batyrev's mirror construction
\cite{batyrev1,bb1}
to (0,2) models defined by certain special (`reflexively plain')
hypersurfaces in
toric varieties, with bundles given by deformations of the tangent bundles.
The authors of \cite{mp-02mirror}
are able to make a proposal for a precise mapping of parameters in
these cases, {\it i.e.} to relate families of (0,2) models, which they
check by matching singularity structures in moduli spaces.

This represents significant progress, but there is still much to do before
(0,2) mirror symmetry is nearly as well understood as ordinary mirror
symmetry.  Beyond the \cite{mp-02mirror} construction, we would still
like a more general mirror construction that applies to a broader class
of varieties, and bundles beyond just deformations of the tangent bundle.
Fully developing (0,2) mirror symmetry will also require further
developments in quantum sheaf cohomology.

\section{Two-dimensional gauge dualities}
\label{sect:duals}

Next, we shall give an overview of recent progress in two-dimensional
gauge theoretic dualities, in which different-looking gauge theories
renormalization-group (RG) flow to the same infrared (IR) fixed point,
{\it i.e.} become isomorphic at low energies and long
distances.

Such dualities are of long-standing interest in the physics
community, and there has been significant recent interest
(see {\it e.g.} \cite{bc1,ggp1,ggp-exact,hori2,jsw,kl2,kl1}).
In two dimensions, we will see that such dualities can at least sometimes
be understood as different presentations of the same geometry.
This not only helps explain why these dualities work, but also implies
a procedure to generate further examples (at least for Calabi-Yau and
Fano geometries).

A prototypical example of a two-dimensional gauge duality, closely
analogous to the central example of four-dimensional Seiberg
duality \cite{seiberg-4d}, was described in \cite{bc1} and relates
a pair of theories with (2,2) supersymmetry:
\begin{center}
\begin{tabular}{cc}
{ \begin{tabular}{c}
$U(k)$ gauge group \\
$n$ chirals in fundamental, $n > k$ \\
$A$ chirals in antifundamental, $A < n$
\end{tabular} } &
{ \begin{tabular}{c}
$U(n-k)$ gauge group \\
$n$ chirals $\Phi$ in fundamental \\
$A$ chirals $P$ in antifundamental \\
$nA$ neutral chirals $M$ \\
superpotential $W = M \Phi P$
\end{tabular} }
\end{tabular}
\end{center}
The theory on the left RG flows to a nonlinear sigma model on
\begin{displaymath}
{\rm Tot}\left( S^{\oplus A} \: \longrightarrow \: G(k,n) \right)
\: = \: \left({\mathbb C}^{kn} \times {\mathbb C}^{kA} \right) //
GL(k) ,
\end{displaymath}
where $S$ is the universal subbundle on the Grassmannian $G(k,n)$.
The RG flow for the theory on the right is a bit more subtle, but can
be analyzed by realizing that the superpotential is realizing a map
in the short exact sequence
\begin{displaymath}
0 \: \longrightarrow \: S \: \stackrel{\Phi}{\longrightarrow} \:
{\mathcal O}^n \: \longrightarrow \: Q \: \longrightarrow \: 0 ,
\end{displaymath}
which implies that the theory on the right RG flows to a nonlinear sigma
model on
\begin{displaymath}
{\rm Tot}\left( (Q^*)^{\oplus A} \: \longrightarrow \: G(n-k,n) \right)
\: = \: 
{\rm Tot}\left( S^{\oplus A} \: \longrightarrow \: G(k,n) \right) ,
\end{displaymath}
the same geometry as for the theory on the left.  Since the two theories
RG flow to nonlinear sigma models on the same geometry, the RG flows
of the two theories eventually coincide, and so the two gauge theories
are Seiberg dual.  In particular, this particular version of Seiberg
duality has a purely geometric understanding.

We can apply the ideas above to make predictions for further two-dimensional
dualities, at least for Fano and Calabi-Yau geometries.
(For other cases, GLSM phases can be decorated with discrete Coulomb
vacua \cite{mp-coulomb,mp-coulomb2}, which complicate the analysis.)

Our next example will be constructed utilizing the fact that
the Grassmannian $G(2,4)$ is a quadric hypersurface in
${\mathbb P}^5$.  The corresponding duality relates the theories
\begin{center}
\begin{tabular}{cc}
{ \begin{tabular}{c}
$U(2)$ gauge theory \\
4 chirals $\phi_i$ in fundamental
\end{tabular} }
&
{ \begin{tabular}{c}
$U(1)$ gauge theory \\
6 chirals $z_{ij} = - z_{ji}$, $i, j =1 \cdots 4$, charge $+1$ \\
one chiral $P$, charge $-2$ \\
$W = P(z_{12} z_{34} - z_{13} z_{24} + z_{14} z_{23} )$
\end{tabular} }
\end{tabular}
\end{center}
The theory on the left RG flows to a nonlinear sigma model on
$G(2,4)$.
The theory on the right RG flows to a nonlinear sigma model on
the corresponding quadric hypersurface.
Since the geometries match, we see that the RG flows converge, and so
the theories are Seiberg dual.  As a consistency check,
the chirals on the right and left are
related by 
\begin{displaymath}
z_{ij} \: = \: \epsilon_{\alpha \beta} \phi_i^{\alpha} 
\phi_j^{\beta} .
\end{displaymath}
Both theories admit a global $GL(4)$ action, which acts as
\begin{displaymath}
\phi_i^{\alpha} \: \mapsto \: V_i^j \phi_j^{\alpha}, \: \: \:
z_{ij} \: \mapsto \: V_i^k V_j^{\ell} z_{k \ell} .
\end{displaymath}
Chiral rings, anomalies, and Higgs moduli spaces match automatically.

This particular example is interesting because
it relates abelian and nonabelian gauge
theories, which in four dimensions would be difficult at best.  In two
dimensions, since gauge fields have no dynamics, abelian and nonabelian
gauge theories are more closely related than in four dimensions.

In two dimensions, this understanding of Seiberg dualities in terms of
matching geometries is not only entertaining but serves a more concrete
purpose.  In four dimensions, renormalizability heavily constrains possible
superpotentials, which means as a practical matter that theories tend to
have a number of global symmetries which can be used as guides to help
confirm possible Seiberg duals.  In two dimensions, by contrast, 
renormalizability does not constrain superpotentials at all, and generic
superpotentials wll break all symmetries.  Thus, identifying gauge duals as
different presentations of the same geometry allows us to construct duals when
standard tricks from four dimensions do not apply.

We can build on the previous example to construct a simple set of 
(2,2) supersymmetric examples in which global symmetries are broken.
Specifically, consider the two theories
\begin{center}
\begin{tabular}{cc}
{ \begin{tabular}{c}
$U(2)$ gauge theory \\
4 chirals $\phi_i$ in fundamental \\
chirals $p_a$, charge $-d_a$ under $\det U(2)$ \\
$W = \sum_a p_a f_a(\epsilon_{\alpha \beta}
\phi_i^{\alpha} \phi_j^{\beta} )$
\end{tabular} }
&
{ \begin{tabular}{c}
$U(1)$ gauge theory \\
6 chirals $z_{ij}=-z_{ji}$ of charge $+1$ \\
one chiral $P$ of charge $-2$ \\
chirals $P_a$ of charge $-d_a$ \\
$W = P(z_{12} z_{34} - z_{13} z_{24} + z_{14} z_{23})$ \\
$\: \: \: \:  \: + \:
\sum_a P_a f_a(z_{ij})$
\end{tabular} }
\end{tabular}
\end{center}
The two theories above RG flow to nonlinear sigma models on the complete
intersection
\begin{displaymath}
G(2,4)[d_1,d_2,\cdots] \: = \:
{\mathbb P}^5[2,d_1,d_2,\cdots]
\end{displaymath}
and so, as above, are Seiberg dual.

An even more complex-appearing (2,2) gauge duality can be described as follows:
\begin{center}
\begin{tabular}{cc}
{ \begin{tabular}{c}
$U(2)$ gauge theory \\
$n$ chirals in fundamental
\end{tabular} }
&
{ \begin{tabular}{c}
$U(n-2)\times U(1)$ gauge theory \\
$n$ chirals $X$ in fundamental of $U(n-2)$ \\
$n$ chirals $P$ in antifundamental of $U(n-2)$ \\
($n$ choose 2) chirals $z_{ij} = - z_{ji}$, \\
$\:\:\:$ charge $+1$ under $U(1)$ \\
$W = {\rm tr}\, PAX$
\end{tabular} }
\end{tabular}
\end{center}
Each of these two theories RG flows to a nonlinear sigma model on $G(2,n)$,
using the fact that $G(2,n)$ can be described as the rank 2 locus of an
$n \times n$ matrix $A$ over ${\mathbb P}^{\frac{n!}{(n-2)! 2!} - 1}$,
where $A$ is defined by
\begin{displaymath}
A(z_{ij}) \: = \: \left[ \begin{array}{cccc}
z_{11} = 0 & z_{12} & z_{13} & \cdots \\
z_{21}=-z_{12} & z_{22}=0 & z_{23} & \cdots \\
z_{31}=-z_{13} & z_{32}=-z_{23} & z_{33}=0 & \cdots \\
\cdots & \cdots & \cdots & \cdots
\end{array} \right],
\end{displaymath}
using the perturbative description of Pfaffians in \cite{hori2,jklmr}.
Since the RG flows converge, the two gauge theories above are necessarily
Seiberg dual.

The same techniques can be extended to two-dimensional theories with (0,2)
supersymmetry.  Consider for example the two theories
\begin{center}
\begin{tabular}{cc}
{ \begin{tabular}{c}
$U(2)$ gauge theory \\
4 chirals in fundamental \\
1 Fermi in $(-4,-4)$ \\
8 Fermis in $(1,1)$ \\
1 chiral in $(-2,-2)$ \\
2 chirals in $(-3,-3)$ \\
plus suitable superpotential
\end{tabular} }
&
{ \begin{tabular}{c}
$U(1)$ gauge theory \\
6 chirals, charge $+1$ \\
2 Fermis, charge $-2, -4$ \\
8 Fermis, charge $+1$ \\
1 chiral, charge $-2$ \\
2 chirals, charge $-3$ \\
plus suitable superpotential
\end{tabular} }
\end{tabular}
\end{center}
(Matter supermultiplets in (0,2) supersymmetry come in two types
labelled `chiral' and `Fermi'.  In the left column, $U(2)$ representations
are indicated with a nonincreasing pair of integers as in \cite{jsw}.)
These theories will RG flow to the (0,2) nonlinear sigma model on the Calabi-Yau
\begin{displaymath}
G(2,4)[4] \: = \: {\mathbb P}^5[2,4] ,
\end{displaymath}
with holomorphic vector bundle ${\mathcal E}$ given as
\begin{displaymath}
0 \: \longrightarrow \: {\mathcal E} \: \longrightarrow \:
\oplus^8 {\mathcal O}(1) \: \longrightarrow \:
{\mathcal O}(2) \oplus^2 {\mathcal O}(3) \: 
\longrightarrow \: 0 .
\end{displaymath}
Since the RG flows converge, these two theories are Seiberg dual.
As a consistency test, it can be shown that the elliptic genera of
these two theories match \cite{jsw}, applying recent GLSM-based computational
methods described in \cite{beht2,beht,gg}.

A different example is provided by `triality' \cite{ggp1,ggp-exact}.
Here, triples of (0,2) GLSMs are believed to flow to the same IR fixed point.
Each GLSM has two different geometric phases; however, unlike previous
cases, not all of the geometric phases describe the same geometry.
Schematically, we can understand the relationship between the phases
as follows \cite{jsw}:
\begin{displaymath}
\xymatrix{
 S^A \oplus (Q^*)^{2k+A-n} \rightarrow G(k,n) 
\ar@{--}[r] \ar@{<->}[d]_{\cong} &
 (S^*)^A \oplus (Q^*)^n \rightarrow G(k,2k+A-n)  
\\
 (Q^*)^A \oplus S^{2k+A-n} \rightarrow G(n-k,n) 
\ar@{--}[r] 
&
 (Q^*)^n \oplus (S^*)^{2k+A-n} \rightarrow G(n-k,A) 
\ar@{<-->}[d]^{\cong} \\
 S^n \oplus (Q^*)^A \rightarrow G(A-n+k,2k+A-n) 
\ar@{--}[r] \ar@{<-->}[d]_{\cong} 
&
 (S^*)^n \oplus (Q^*)^{2k+A-n} \rightarrow G(A-n+k,A)  \\
 (Q^*)^n \oplus (S^*)^A \rightarrow G(k,2k+A-n) 
\ar@{--}[r] 
& 
(Q^*)^{2k+A-n} \oplus S^A \rightarrow G(k,n) .
}
\end{displaymath}
Each phase also has a bundle summand, either $(\det S)^{\oplus 2}$ or
$(\det S^*)^{\oplus 2}$, which we have omitted for brevity.
Horizontal dashed lines indicate phase transitions to different
geometries; vertical arrows indicate equivalent geometries.
The fourth line is physically equivalent to the first:  the bottom right
corner is equivalent to the upper left, and the bottom left, to the
upper right.
In writing the diagram above, we have used the fact that in (0,2) theories,
dualizing the gauge bundle is an equivalence of the theories:
\begin{displaymath}
{\rm QFT}(X, {\mathcal E} \rightarrow X) \: \cong \:
{\rm QFT}(X, {\mathcal E}^* \rightarrow X) .
\end{displaymath}
(See for example \cite{es-other} for a discussion of corner cases of
this duality.)  A test of triality recently appeared in
\cite{gjs}.

How do these gauge dualities relate to (0,2) mirrors as discussed in the
previous section?  As we have seen, gauge dualities often relate different
presentations of the same geometry, whereas (0,2) mirrors exchange different
geometries.  The existence of (0,2) mirrors seems to imply that there ought to
exist more `exotic' gauge dualities, that present different geometries.

So far in this section
we have used mathematics to make predictions for physics.
In the next section we shall turn that around, and use physics to make
predictions for mathematics.

\section{Decomposition in two-dimensional nonabelian gauge theories}
\label{sect:decomp}

In a two-dimensional orbifold or gauge theory, if a finite subgroup of
the gauge group acts trivially on all massless matter, the theory decomposes
as a disjoint union \cite{hhpsa}.

For example, a trivially-acting ${\mathbb Z}_2$ orbifold of a nonlinear
sigma model on a space $X$
is equivalent to a nonlinear sigma model on two copies of $X$:
\begin{displaymath}
{\rm CFT}\left( [X/{\mathbb Z}_2 ] \right) \: = \:
{\rm CFT}\left( X \coprod X \right) .
\end{displaymath}
In the ${\mathbb Z}_2$ orbifold, since the ${\mathbb Z}_2$ acts
trivially on $X$, there is a dimension zero twist field.  Linear combinations
of that twist field and the identity operator form projection operators 
onto the two copies of $X$.

For another example, consider a $D_4$ orbifold of a nonlinear sigma model
on a space $X$, where the center ${\mathbb Z}_2 \subset D_4$ acts
trivially on $X$.  This orbifold is equivalent to the disjoint union of
a pair of ${\mathbb Z}_2 \times {\mathbb Z}_2$ orbifolds, one with and
one without discrete torsion:
\begin{displaymath}
{\rm CFT}\left( [X/D_4] \right) \: = \:
{\rm CFT}\left( [X/{\mathbb Z}_2 \times {\mathbb Z}_2] \coprod
[X/{\mathbb Z}_2 \times {\mathbb Z}_2]_{\rm d.t.}\right) ,
\end{displaymath}
where $D_4/{\mathbb Z}_2 = {\mathbb Z}_2 \times {\mathbb Z}_2$.

These are examples in physics of what is meant by `decomposition.'

Decomposition is also a statement about mathematics.  Briefly, over the last
several years, the following dictionary has been built:
\begin{center}
\begin{tabular}{cc}
$\,$ & \\
2d Physics & Math \\ \hline
D-brane & Derived category \cite{medc} \\
Gauge theory & Stack \cite{glsm,msx,nr} \\
Gauge theory with & Gerbe \cite{hhpsa,glsm,msx,nr} \\
trivially-acting subgroup & \\
$\,$ & \\
Universality class of & Categorical equivalence \\
renormalization group flow &  \\
$\,$ & 
\end{tabular}
\end{center}
In particular, decomposition is a statement about the physics of strings
propagating on gerbes, detailed in the `decomposition conjecture' \cite{hhpsa},
which for banded gerbes can be summarized as:
\begin{displaymath}
{\rm CFT}\left( G-\mbox{gerbe on }X \right) \: = \:
{\rm CFT}\left( \coprod_{\hat{G}} (X, B) \right) ,
\end{displaymath}
where $\hat{G}$ is the set of irreducible representations of $G$,
and the $B$ field on each component is determined by the image of
the characteristic class of the gerbe:
\begin{displaymath}
H^2(X,Z(G)) \: \stackrel{ Z(G) \mapsto U(1) }{\longrightarrow} \:
H^2(X,U(1)) .
\end{displaymath}
The decomposition conjecture has been checked in a wide variety of ways,
including, for example:
\begin{itemize}
\item multiloop orbifold partition functions:  partition functions
decompose in the desired form,
\item quantum cohomology ring relations as derived from GLSMs match the
implicit prediction above,
\item D-branes, K theory, sheaves on gerbes:  the physical decomposition
of D-branes matches the mathematical decomposition of K theory and sheaves
on gerbes.
\end{itemize}

Decomposition also has a number of applications, including
\begin{itemize}
\item Predictions for Gromov-Witten invariants of gerbes,
as checked in {\it e.g.} \cite{ajt1,ajt2,ajt3,ajt4,gt1,tt1,tseng-deg0,tseng1},
\item Understanding certain GLSM phases 
\cite{cdhps,hkm,hori2,es-rflat}, 
via giving a physical
realization of Kuznetsov's homological projective duality 
\cite{kuz2},
\end{itemize}
and these works serve implicitly as further checks on the
decomposition conjecture above.

To understand the decomposition conjecture in orbifolds, one can compare
(multi)loop partition functions, state spaces, and D-branes, and they
all imply the same result.  In gauge theories, there are further subtleties.
For example, let us compare the following two theories:
\begin{itemize}
\item Ordinary ${\mathbb C}{\mathbb P}^n$ model:  a $U(1)$ gauge theory with
$n+1$ chiral superfields, each of charge $+1$,
\item Gerby ${\mathbb C}{\mathbb P}^n$ model:  a $U(1)$ gauge theory with
$n+1$ chiral superfields, each of charge $k$, $k > 1$.
\end{itemize}
In order for these two theories to be distinct, the physics of the second
must be different from the first -- but how can multiplying the charges
by a factor change anything?  Naively, this is just a convention, and
physics should not depend upon conventions.

Perturbatively, multiplying all the charges by a factor does not modify
the physics; however, nonperturbatively\footnote{
We would like to thank A.~Adams, J.~Distler, and R.~Plesser for explaining
the distinction, on both compact and noncompact worldsheets, at an Aspen
workshop in 2004.
}, there can be a difference between
these two theories.  On a compact worldsheet, to make manifest the
distinction, one must specify which bundles the fields couple to, to 
unambiguously specify the theory.  If the chiral fields are sections of
a line bundle $L$ in the first theory, then in the second they are
sections of a different bundle, $L^{\otimes k}$, and hence have different
zero modes, different anomalies, and hence different nonperturbative
physics.

On a noncompact worldsheet, one can instead appeal to the periodicity of the
$\theta$ angle in the two-dimensional gauge theory.  The $\theta$ angle
acts as an electric field, so by building a sufficiently large capacitor,
one can excite states of arbitary mass.  In particular, we can distinguish
the second theory from the first by adding a pair of massive minimally
charged fields, which a sufficiently large capacitor can excite.
In this fashion, essentially through different periodicities of the
$\theta$ angle, one can distinguish the two theories.

Now, decomposition has been extensively checked for orbifolds and
abelian gauge theories, but tests in nonabelian gauge theories in
two dimensions have only appeared more recently
\cite{es-nonabelian-decomp}.  Since two-dimensional gauge theories do not
have propagating degrees of freedom, an analogous phenomena ought to take
place in nonabelian gauge theories with center-invariant matter.
Specifically, it was proposed in \cite{es-nonabelian-decomp} that
for $G$ semisimple, a $G$-gauge theory with center-invariant matter
should decompose into a sum of theories with variable discrete theta
angles.  For example, an $SU(2)$ gauge theory with only adjoints or
other center invariant matter should decompose into a pair of $SO(3)$
gauge theories with the same matter but different discrete theta angles,
schematically:
\begin{equation}   \label{eq:su2-decomp}
SU(2) \: = \: SO(3)_+ \: + \: SO(3)_- .
\end{equation}

Before working through this in detail, let us first remind the reader of
how discrete theta angles are defined, as they are relatively
new \cite{ast,gmn}.
Consider a two-dimensional gauge theory, with gauge group $G = \tilde{G}/K$,
$\tilde{G}$ compact, semisimple, and simply-connected, $K$ a finite subgroup
of the center of $\tilde{G}$.  This theory has a degree-two $K$-valued
characteristic class which we will denote $w$.  (For example, in an
$SO(3)$ gauge theory, this is the second Stiefel-Whitney class.)
For any character
$\lambda$ of $K$, we can add the topological term $\lambda(w)$
to the action.  This is the discrete theta angle term, and we see in
this fashion that the possible values of the discrete theta angle are
classified by characters of $K$.

For example, let us consider an $SO(3)$ gauge theory.  Now,
\begin{displaymath}
SO(3) \: = \: SU(2) / {\mathbb Z}_2 ,
\end{displaymath}
hence as ${\mathbb Z}_2$ has two characters, we see that an $SO(3)$ gauge
theory in two dimensions has two discrete theta angles.

Let us check the decomposition conjecture for nonabelian gauge theories
in the case of pure $SU(2)$ gauge theory in two dimensions.
The partition function for pure (nonsupersymmetric) two-dimensional
gauge theories can be found in {\it e.g.} \cite{grosstaylor,migdal1,rusakov1},
from which we derive
\begin{eqnarray*}
Z(SU(2)) & = & \sum_R (\dim R)^{2-2g} \exp(-A C_2(R)), \\
Z(SO(3)_+) & = & \sum_R (\dim R)^{2-2g} \exp(-A C_2(R)).
\end{eqnarray*}
In the expressions above, $g$ is the genus of the two-dimensional surface,
$A$ is its area, $R$ a representation, and $C_2(R)$ a Casimir of the
representation $R$.  The $SU(2)$ partition function sums over all
representations $R$ of $SU(2)$, and the $SO(3)_+$ partition function
sums over all representations $R$ of $SO(3)$.  (For $SO(3)_+$, the
discrete theta angle vanishes, so $SO(3)_+$ is the ordinary $SO(3)$
gauge theory.)  The partition function of $SO(3)_-$ was described
in \cite{tachikawa1}, and has the form
\begin{displaymath}
Z(SO(3)_-) \: = \: \sum_R (\dim R)^{2-2g} \exp(-A C_2(R)) ,
\end{displaymath}
where the sum is now over representations of $SU(2)$ that are not
representations of $SO(3)$.  Combining these three expressions, it should
be clear that
\begin{displaymath}
Z(SU(2)) \: = \: Z(SO(3)_+) \: + \: Z(SO(3)_-) .
\end{displaymath}

More generally, for $G$ gauge theories with $G$
semisimple, $K$ a finite subgroup of the center of
$G$, and matter invariant under $K$,
we can express decomposition schematically as
\begin{displaymath}
G \: = \: \sum_{\lambda \in \hat{K}} (G/K)_{\lambda} .
\end{displaymath}
This can be checked for pure gauge theories using partition functions
as above, and can also similarly be checked for correlation functions
of Wilson lines in pure gauge theories.  
In addition, it can also be checked in supersymmetric theories
using expressions for partition functions given in 
\cite{bc1,doroudetal}. The arguments in this case revolve around details
of cocharacter lattices, which for brevity we omit here; 
see \cite{es-nonabelian-decomp} for details.

\section{Heterotic moduli}
\label{sect:moduli}

It was known historically that for large-radius heterotic nonlinear
sigma models on the (2,2) locus, there were three classes of infinitesimal
moduli\footnote{
Physically, the moduli are indistinguishable from one another; the
distinction we list is purely mathematical in origin.
}:
\begin{itemize}
\item K\"ahler moduli, counted by $H^1(X, T^*X)$, 
\item Complex moduli, counted by $H^1(X, TX)$,
\item Bundle moduli, counted by $H^1(X, {\rm End}\, {\mathcal E})$,
\end{itemize}
for a compactification on a space $X$ with gauge bundle 
${\mathcal E} = TX$ (the (2,2) locus).  

When the gauge bundle ${\mathcal E}$ is different
from the tangent bundle $TX$, the correct counting is more complicated.
It was shown in the physics literature in
{\it e.g.} \cite{aglo1} that the correct counting is given by
\begin{itemize}
\item K\"ahler moduli, counted by $H^1(X, T^*X)$, 
\item Compatible complex and bundle moduli, counted by $H^1(Q)$ where
$Q$ is defined by the Atiyah sequence
\begin{equation}  \label{eq:atiyah}
0 \: \longrightarrow \: {\rm End }\,{\mathcal E} \: \longrightarrow \:
Q \: \longrightarrow \: TX \: \longrightarrow \: 0 .
\end{equation}
The extension class is determined by the curvature of the bundle.
Specifically, it is an element of
\begin{displaymath}
{\rm Ext}^1(TX, {\rm End}\, {\mathcal E}) \: = \: H^1(T^*X \otimes
{\rm End}\, {\mathcal E})
\end{displaymath}
given by the curvature.
\end{itemize}
In particular, as ${\mathcal E}$ is required to be a {\it holomorphic}
bundle, the complex and bundle moduli are not independent of one another,
and in fact a given bundle may not be compatible with all complex
structure moduli, a result that was well-known in mathematics but
whose relevance the physics community only recently digested.

At the time, however, this still left unresolved the question of 
understanding moduli of heterotic non-K\"ahler compactifications
\cite{strom1}.  In a non-K\"ahler compactification, there is no
version of Yau's theorem relating metric moduli to complex and K\"ahler
moduli, so in principle, in close-to-large-radius\footnote{
Non-K\"ahler heterotic compactifications do not have a large-radius limit.
The best one can do is to hope for solutions
``close'' to large-radius, where geometry is still valid.  In this section,
we implicitly assume the non-K\"ahler compactifications being considered
are all in that regime, close enough to large radius that geometry is a valid
description.
} non-K\"ahler 
compactifications, the moduli need not have any meaningful connection
to Calabi-Yau moduli.  (That said, it should also be noted that even in
a Calabi-Yau (0,2) compactification, although the space admits a K\"ahler
metric, away from the large-radius limit
the metric solving the supergravity equations is necessarily non-K\"ahler, 
because the Green-Schwarz condition forces $H$ to be nonzero.)

A partial solution to this problem was discovered
in \cite{ms-nk}.  There, it was argued from a worldsheet
analysis that for non-K\"ahler
compactifications in a purely formal $\alpha' \rightarrow 0$ limit,
the infinitesimal moduli are counted by $H^1(S)$, where
\begin{displaymath}
0 \: \longrightarrow \: T^*X \: \longrightarrow S \: \longrightarrow \:
Q \: \longrightarrow \: 0,
\end{displaymath}
where $Q$ is the extension determined by the Atiyah sequence~(\ref{eq:atiyah}).
The extension above is determined by an element of
\begin{displaymath}
{\rm Ext}^1(TX,T^*X)
\end{displaymath}
determined by the $H$ flux, which is assumed to obey $d H = 0$.

As non-K\"ahler compactifications do not exist in the $\alpha'
\rightarrow 0$ limit, the solution above was necessarily incomplete.
It was improved upon in \cite{ags,delaOssa:2014cia}, 
which gave an overcounting of heterotic
moduli valid through first order in $\alpha'$.
On manifolds satisfying the $\partial \overline{\partial}$-lemma,
the moduli are overcounted by $H^1(S')$, where
\begin{displaymath}
0 \: \longrightarrow \: T^*X \: \longrightarrow \: S' \: \longrightarrow
\: Q' \: \longrightarrow \: 0
\end{displaymath}
(defined by $H$ satisfying the Green-Schwarz condition), for $Q'$ given by
\begin{displaymath}
0 \: \longrightarrow \: {\rm End} \, {\mathcal E} \oplus 
{\rm End} TX \: \longrightarrow \: Q' \: \longrightarrow \: TX
\: \longrightarrow \: 0 ,
\end{displaymath}
with the extension defined by the curvatures of the gauge bundle and $TX$.

The overcounting above is the current state-of-the-art; currently
work is in progress to find the correct counting and to extend to higher
orders in $\alpha'$.

So far we have outlined infinitesimal moduli, corresponding to
marginal operators on the worldsheet.  These can be obstructed by
{\it e.g.} nonperturbative effects, and there is an interesting story
behind this.  Initially, in the mid-80s, it was observed in
\cite{dsww1,dsww2} 
that a single worldsheet instanton can generate a superpotential
term obstructing deformations off the (2,2) locus, but in the early 90s
it was observed that for moduli realizable in GLSMs, the sum of the
contributions from different contributing rational curves all cancel
out, and so the
moduli are unobstructed.  This led to a revitalization of interest in
(0,2) models, and paved the way for work on F theory, for example.
The original GLSM arguments have found alternate presentations\footnote{
In our experience, sometimes these papers are mis-quoted as claiming that
the spacetime superpotential vanishes in heterotic compactifications.
The correct statement is that nonperturbative corrections to
gauge singlet moduli interactions arising from moduli realizable in GLSMs
cancel out.  However, gauge non-singlet interactions can and will receive
nonperturbative corrections, and can even be nonzero classically.  
For example, on the (2,2) locus,
in a heterotic compactification on a Calabi-Yau 3-fold, the 
${\bf \overline{27}}^3$ couplings are nonzero:  in addition to the
classical contribution described in \cite{strom-yuk},
they also receive nonperturbative
corrections corresponding to the Gromov-Witten invariants of the Calabi-Yau
\cite{candelas-mirror,witten-tft}.
} in
{\it e.g.} \cite{basu-sethi,beasley-ed}.  Current work on the subject,
such as \cite{paul-ben,elephants,paul-ronen,bkos1,bkos4,bkos3,bkos2}, 
has focused on understanding
non-GLSM moduli, for which nonperturbative corrections to
obstructions often do not cancel out.

\section{Conclusions}

In this note we have given an overview of recent developments in
two-dimensional theories, focusing primarily though not exclusively
on (0,2) theories.
We began in section~\ref{sect:qsc} with a brief review of the current state of
the art in quantum sheaf cohomology.  In section~\ref{sect:02mirror} we
gave a brief status report on (0,2) mirror symmetry.
In section~\ref{sect:duals} we discussed recent progress in
two-dimensional gauge dualities in theories with (2,2) and (0,2)
supersymmetry.  We showed how a number of Seiberg-like dualities can
be understood simply as different presentations of the same IR geometry,
and use this to predict additional dualities.
In section~\ref{sect:decomp} we described a different gauge duality,
one that applies to both supersymmetric and nonsupersymmetric theories in
two dimensions.  Specifically, in two-dimensional gauge theories in which
a finite subgroup of the gauge group acts trivially on the matter, the
theory `decomposes' into a disjoint union of theories.  In nonabelian gauge
theories, the various components are labelled by different discrete theta
angles.
Finally, in section~\ref{sect:moduli} we discussed current progress in
infinitesimal moduli in heterotic compactifications, specifically,
recent developments in understanding moduli
in both Calabi-Yau and also non-K\"ahler heterotic compactifications.

\bibliographystyle{amsplain}
\bibliography{article}

\end{document}